\documentclass{article}
\usepackage{spconf,amsmath,amssymb,graphicx,bbm,acronym}
\usepackage[inline]{enumitem}

\def\x{{\mathbf x}}
\def\X{{\mathbf X}}
\def\Y{{\mathbf Y}}
\def\xt1{{\mathbf x}_{\left[t-1\right]}}

\newacro{LSTM}[LSTM]{Long Short Term Memory}
\newacro{DNN}[DNN]{Deep Neural Network}
\newacro{CRNN}[CRNN]{Convolutional Recurrent Neural Network}
\newacro{RNN}[RNN]{Recurrent Neural Network}
\newacro{GRU}[GRU]{Gated Recurrent Unit}
\newacro{AHC}[AHC]{Agglomerative Hierachical Clustering}
\newacro{PLDA}[PLDA]{Probabilistic Linear Discriminat Analysis}
\newacro{DER}[DER]{Diarization Error Rate}
\newacro{SAD}[SAD]{Speech Activity Detection}
\newacro{PCA}[PCA]{Principal Component Analysis}
\newacro{MSE}[MSE]{Mean Squared Error}
\newacro{SML}[SML]{Sample Mean Loss}
\newacro{ddCRP}[ddCRP]{distance dependent Chinese Restaurant Processes}

\title{Supervised online diarization with sample mean loss \\ for multi-domain data}
\name{Enrico Fini$^1$, Alessio Brutti$^2$}
\address{$^1$PerVoice Spa, Trento (Italy), $^2$Fondazione Bruno Kessler, Trento (Italy)\\
\small{\tt{enrico.fini@gmail.com, brutti@fbk.eu}}}{}{}

\begin{document}
\ninept
\maketitle
\begin{abstract}
Recently, a fully supervised speaker diarization approach was proposed (UIS-RNN) which models speakers using multiple instances of a parameter-sharing recurrent neural network. In this paper we propose qualitative modifications to the model that significantly improve the learning efficiency and the overall diarization performance. In particular, we introduce a novel loss function, we called \textit{Sample Mean Loss} and we present a better modelling of the speaker turn behaviour, by devising an analytical expression to compute the probability of a new speaker joining the conversation. In addition, we demonstrate that our model can be trained on fixed-length speech segments, removing the need for speaker change information in inference. Using x-vectors as input features, we evaluate our proposed approach on the multi-domain dataset employed in the DIHARD II challenge: our online method improves with respect to the original UIS-RNN and achieves similar performance to an offline agglomerative clustering baseline using PLDA scoring. 


\end{abstract}
\begin{keywords}
Speaker diarization, x-vectors, clustering, supervised learning, recurrent neural networks
\end{keywords}
\section{Introduction}
\label{sec:intro}
The speaker diarization task consists in establishing ``who spoke when'' in a given audio recording~\cite{tranter2006,Anguera2012overview}. Despite having been investigated for decades, diarization is still an unsolved problem in many real scenarios, as highlighted by the recent DIHARD I and DIHARD II challenges~\cite{Ryant2019}.

Typically, speaker diarization is addressed integrating several different components: voice activity detection, speaker change detection, feature extraction and clustering. Most of the research works in literature focus on extracting highly discriminative feature vectors. The first example in this direction are i-vectors~\cite{kenny2008,dehak2011}, which represent a given utterance with a single fixed-dimensional feature vector. The recent rise of neural paradigms has led to the introduction of a variety of approaches to extract the so-called speaker embeddings. These are, typically, derived from the outputs of the inner layers of a neural network trained on a speaker classification task~\cite{Nagrani18}. The most popular embeddings are d-vectors~\cite{variani2014deep} and x-vectors \cite{garcia2017speaker, Sell2018}.

Conversely, not much progress has been done with regard to clustering. In most of the approaches, this stage is still based on the~\ac{AHC}~\cite{han2008strategies} in combination with~\ac{PLDA} scoring~\cite{sell2014speaker}. Recently, spectral clustering~\cite{ning2006spectral}\cite{wang2018speaker}, and variatioanl bayesian clustering~\cite{Diez_2018,Diez2019} have been introduced, showing promising result. Also, alternatives to the~\ac{PLDA} scoring have been introduced using neural networks that learn how to score two speech segments~\cite{Chopra2005}, using siamese networks~\cite{zhu2018self} or Bi-LSTMs~\cite{Lin2019}. Nevertheless, clustering remains unsupervised and heavily dependent on fine-tuned hyperparameters (e.g. thresholds to stop clustering).

Recently, efforts have been made to formulate clustering in a supervised learning framework~\cite{Asawa2017DeepLA,Fujita_2019,zhang2019fully}. Supervised clustering is attractive because it can be optimized on the diarization metrics directly, or learning context dependent parameters. Additionally, supervision allows to improve performance by learning from the increasing amount of data at our disposal. For example,~\cite{Asawa2017DeepLA} tackles the diarization problem as a classification task, while~\cite{Fujita_2019} uses a permutation invariant loss and a clustering loss to dynamically identify speakers. Both \cite{Asawa2017DeepLA} and \cite{Fujita_2019} assume that the number of speaker is known apriori or at least bounded. This assumption is removed in the UIS-RNN~\cite{zhang2019fully}: a fully supervised approach which handles an unbound number of speakers using an online generative process. Speaker distributions are modelled with multiple instances of a parameter-sharing~\ac{RNN}. A further, strong advantage of~\cite{zhang2019fully} over traditional clustering algorithms is the fact that decoding is online using beam search~\cite{medress1977}. Though online diarization had already been explored, using both unsupervised \cite{geiger2010gmm,mansfield2018links,Dimitriadis2017DevelopingOS} and supervised \cite{Asawa2017DeepLA} paradigms, the UIS-RNN stands out in terms of performance, 
outperforming the previous offline state of the art on telephone data.

Although these are very interesting results, an online system that works well across multiple domains still remains an open problem. As a matter of fact, diarization systems presented in the literature appear to
work relatively well on domains with a low number of speakers and no overlapping speech, like telephone data, while performance tends to deteriorate in more challenging contexts such as meetings or dinner parties.


In this paper we present an evolution of the UIS-RNN~\cite{zhang2019fully}, which substantially improves the performance. First of all, we introduce a new loss function for training the~\ac{RNN} that models speakers, which provides faster convergence, encouraging the network to find deeper minima, and generalizes better on the evaluation set. Secondly, we propose a semantically grounded formulation for the unseen speaker intervention probability that is easy to calculate and improves performance in inference. In addition we train on fixed-length speech segments, and let the neural network aggregate embeddings,  removing the constraint on speaker change information in inference. Finally, we shed light on the performance of the proposed method with respect to the original UIS-RNN in a multi-domain scenario through extensive testing on the DIHARD datasets. We also make our results reproducible, since we use a publicly available embedding extractor and fully disclose our code\footnote{The first author performed this work as an intern at PerVoice and Fondazione Bruno Kessler. The implementation of this paper is available at:
\tt{https://github.com/DonkeyShot21/uis-rnn-sml}}.



\section{Proposed approach}
\label{sec:proposedapproach}

Given a set of embeddings $\X=\left(\x_1,\dots,\x_T\right)$ 
and the related speaker labels $\Y=\left(y_1,\dots,y_T\right)$, where $T$ is the total number of observations, we can cast the diarization problem in a probabilistic framework, looking for the sequence of speaker labels that maximizes the joint probability:
\begin{equation}
\label{eq:onlinegenerative}
    \hat{\Y}=\arg\max_{\Y}P(\X,\Y),
\end{equation}
If we model eq.~\ref{eq:onlinegenerative} as an online generative problem as in~\cite{zhang2019fully}, we can rewrite the joint probability at time $t$ as:
\begin{multline}
\label{eq:factorized}
p(\x_t,y_t,z_t|\xt1,y_{\left[t-1\right]},z_{\left[t-1\right]})\\
=\underbrace{p(\x_t|\xt1,y_t)}_\text{sequence generation}\cdot \underbrace{p(y_t|y_{\left[t-1\right]},z_t)}_\text{assignment} \cdot \underbrace{p(z_t|z_{\left[t-1\right]})}_\text{speaker change},
\end{multline}
where $z_{t}=\mathbbm{1}\left(y_{t} \neq y_{t-1}\right)$ is a hidden binary indicator of speaker change and $\left[t\right]$ denotes all observations up to $t$ included. In the original definition of the UIS-RNN~\cite{zhang2019fully}, the {\bf speaker change} term of eq.~\ref{eq:factorized} is modelled by a coin flipping process where the only parameter is $p_0$, the transition probability. The {\bf speaker assignment} term is implemented as a \ac{ddCRP}~\cite{blei2011distance}, a Bayesian nonparametric process that guides how speakers interleave in the time domain.
Finally, the {\bf sequence generation} part of eq.~\ref{eq:factorized} is modelled using an \ac{RNN}, specifically a \ac{GRU}, that parametrizes the distribution of embeddings assuming a Gaussian distribution as follows:
\begin{equation}
\label{eq:gaussian}
    \x_t|\x_{[t-1]}, y_{[t]}\sim \mathcal{N}\left(\boldsymbol{\mu}\left(GRU_{\boldsymbol{\theta}}\left( \x_{t'}\in\x_{[t-1]} \middle| y_{t'} = y_t\right)\right),{\mathbf \sigma^2\boldsymbol{I}}\right),
\end{equation}
where $\boldsymbol{\mu}\left(GRU_{\boldsymbol{\theta}}\left(\boldsymbol{\cdot}\right) \right)$ is the averaged output of the neural network with parameters $\boldsymbol{\theta}$ instantiated for speaker $y_t$.

\subsection{Original UIS-RNN training}
\begin{figure}[!th]
\centering
  \includegraphics[width=0.75\columnwidth]{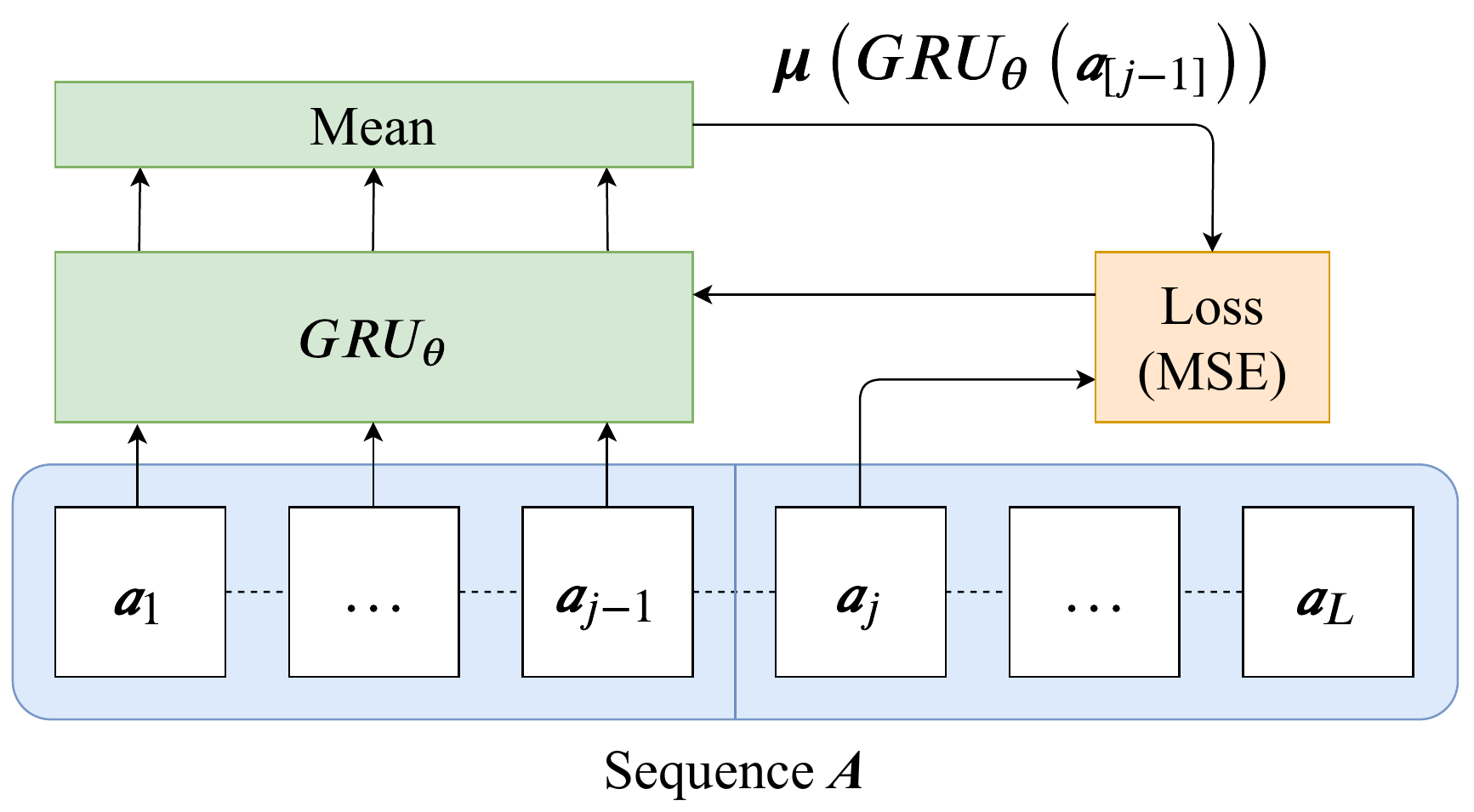}
  \caption{Block diagram of the original UIS-RNN training strategy for a generic sequence $\boldsymbol{A}$.}
  \label{fig:stock_boxes}
\end{figure}

Given a dataset $\boldsymbol{\mathcal{D}} = \{ \left(\boldsymbol{\X}_{1}, \dots, \boldsymbol{\X}_{M} \right), \left(\boldsymbol{\Y}_{1}, \dots, \boldsymbol{\Y}_{M} \right)\}$, including $M$ sequences of embeddings and related label, the optimal set of network parameters $\boldsymbol{\theta}^{*}$ can be obtained minimizing the following negative log likelihood \cite{zhang2019fully}: 
\begin{equation}
    \label{eq:nll}
    \mathcal{L} = \sum_{m=1}^{\left|\boldsymbol{\mathcal{D}}\right|} - \ln{p(\X_m | \Y_m; \boldsymbol{\theta})}.
\end{equation}
Using the model in eq.~\ref{eq:gaussian}, eq.~\ref{eq:nll} can be reformulated in a~\ac{MSE} fashion~\cite{uisrnn-official-library}:
\begin{equation}
\label{eq:networktraining}
    \mathcal{L}_{\mbox{\tiny{MSE}}} =  \sum_{i=1}^{|\boldsymbol{\mathcal{D}}_A|} \sum_{j=1}^{|\boldsymbol{A}_i|}{\left\| \boldsymbol{a}_{i,j} -\boldsymbol{\mu}\left(GRU_{\boldsymbol{\theta}}\left(\boldsymbol{a}_{i,[j-1]}\right)\right) \right\|}^2.
\end{equation}
Given $S$ speakers and $P$ permutations applied to the data for augmentation purposes, $\boldsymbol{\mathcal{D}}_A = \left(\boldsymbol{A}_{1}, \dots, \boldsymbol{A}_{S \times P} \right)$ is a set of single speaker sequences, where each sequence $\boldsymbol{A}_{i} = (\boldsymbol{a}_{i,1}, \dots,\boldsymbol{a}_{i,L_i}) \in \boldsymbol{\mathcal{D}}_A$ is obtained by concatenating a random permutation of the embeddings generated by the $i$-th speaker. $L_i$ and $\boldsymbol{a}_{i,j}$ are respectively the length and the $j$-th embedding of sequence $\boldsymbol{A}_i$. 

Note that, since the sequences are shuffled, the network can not learn any causal relationship between observations and how to predict the next embedding. Basically, the network is trained to generate samples $\boldsymbol{\varkappa}_t$ from an auxiliary distribution $q$ with expectation $\mathbb{E}[\boldsymbol{\varkappa}_t]$ equal to $\mathbb{E}[\x_t]$. Therefore, the network will learn to predict the mean of the distribution of the embeddings.
%
Figure~\ref{fig:stock_boxes} graphically describes the training presented in~\cite{zhang2019fully} and implemented in~\cite{uisrnn-official-library}.

\subsection{Sample Mean Loss training: UIS-RNN-SML}

\begin{figure}[!ht]
\centering
  \includegraphics[width=0.75\columnwidth]{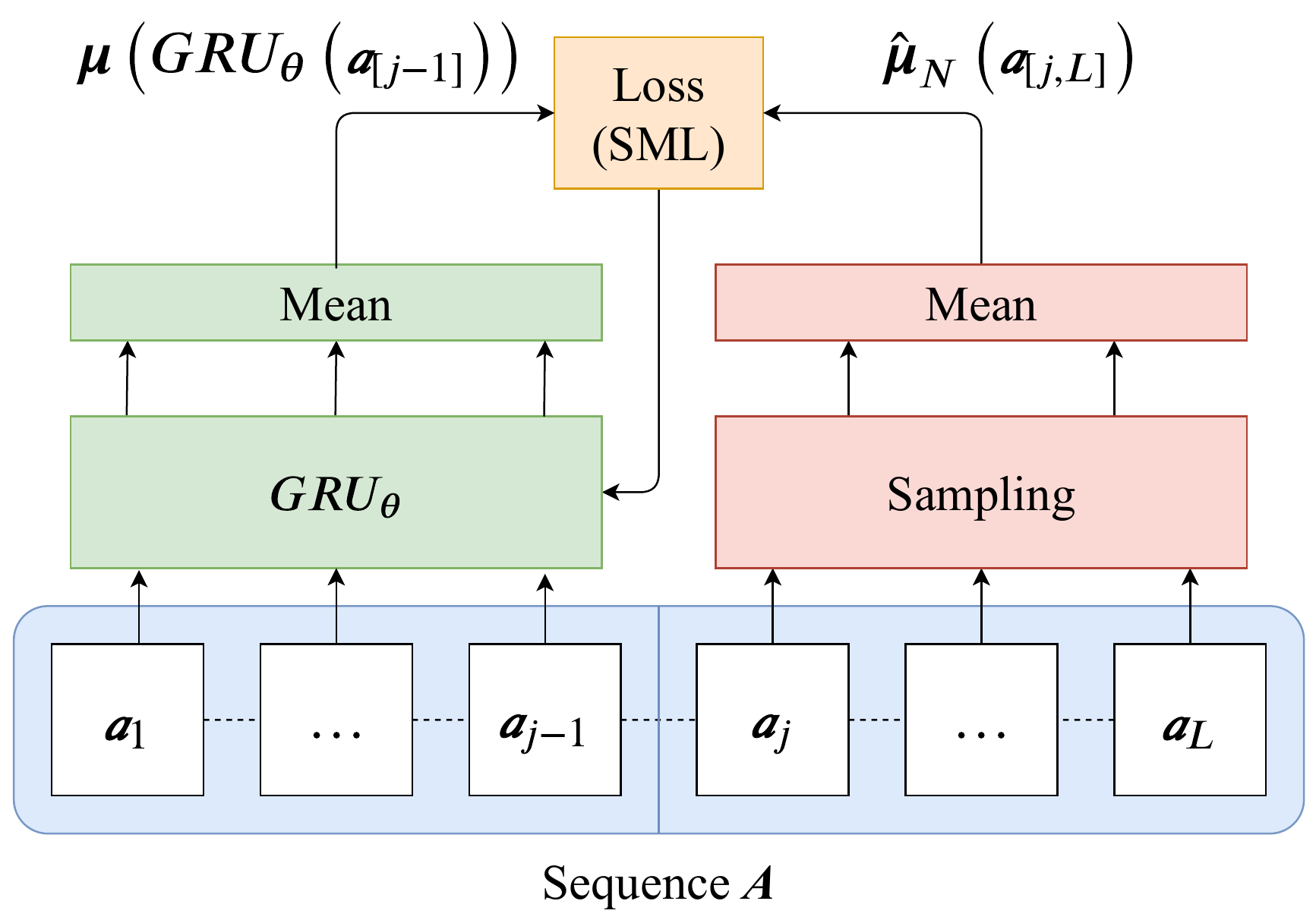}
  \caption{Block diagram of the proposed UIS-RNN-SML training approach for a generic sequence $\boldsymbol{A}$.}
  \label{fig:ours_flow}
\end{figure}

In this section we propose a modified loss that relies on more accurate targets for the network output. Rather than adjusting the network by comparing the mean of its outputs with the next observed embedding, we define a~\ac{MSE} loss with respect to the actual mean of the speaker embeddings of a given speaker. This results in defining a predictor of the mean of the embedding distribution, having seen only a small sample of it. 
More formally, we replace the~\ac{MSE} loss in eq.~\ref{eq:networktraining} with: 
%
\begin{equation}
\label{eq:mean}
\mathcal{L}' =  \sum_{i=1}^{|\boldsymbol{\mathcal{D}}_A|} \sum_{j=1}^{|\boldsymbol{A}_i|} {\left\| \mathbb{E}\left[s\left(i\right) \right]  -\boldsymbol{\mu}\left(GRU_{\boldsymbol{\theta}}\left(\boldsymbol{a}_{i,[j-1]}\right)\right) \right\|}^2,
\end{equation}
where $s\left(i\right)$ is a function that maps each index $i$ to the embedding distribution of the speaker who generated the observations in $\boldsymbol{A}_i$. 

In practice, the actual probability distribution of the embeddings is not available. In addition, given the limited amount of labelled data, using the bare mean over the sequence would lead to overfitting. Therefore, we build the ground truth for the network by estimating the mean over a collection of unseen samples we draw randomly with replacement from the permuted sequence itself. In formulas, given a generic sequence $\boldsymbol{A}_i$ and a subset $\boldsymbol{H} = \left(\boldsymbol{h}_1,\dots,\boldsymbol{h}_N \right) \subseteq\boldsymbol{A}_i$ of $N$ randomly sampled embeddings, we estimate the mean of the embeddings as:
$\hat{\boldsymbol{\mu}}_N\left(\boldsymbol{A}_i\right) = \left({\sum_{i}^N {\boldsymbol{h}_i}}\right)/{N}$
Eq.~\ref{eq:mean} is then rewritten leading to our~\ac{SML} definition:
\begin{equation}
\label{eq:sample-mean}
\mathcal{L}_{\mbox{\tiny{SML}}} = \sum_{i=1}^{|\boldsymbol{\mathcal{D}}_A|} \sum_{j=1}^{|\boldsymbol{A}_i|} {\left\| \hat{\boldsymbol{\mu}}_N\left(\boldsymbol{a}_{i,[j,L_i]}\right) -\boldsymbol{\mu}\left(GRU_{\boldsymbol{\theta}}\left(\boldsymbol{a}_{i,[j-1]}\right)\right) \right\|}^2,
\end{equation}
where we denote the ordered set $(j,\dots,L_i)$ as $[j,L_i]$. Figure~\ref{fig:ours_flow} depicts the proposed training approach for a generic sequence $\boldsymbol{A}$.


\subsection{New speaker probability}
\label{sec:crp_alpha}
One of the most interesting advantages of the UIS-RNN~\cite{zhang2019fully} over other supervised methods, like~\cite{Fujita_2019}, is its ability to model an unbounded number of speakers. This is achieved using a \ac{ddCRP} model~\cite{blei2011distance} that provides the probability of switching back to a previously seen speaker proportionally to the number of turns of that speaker and accounts for the probability of a new speaker joining the conversation. Assuming speakers are numerated in order of appearance starting from $1$, we let:
\begin{align} p\left(y_{t}=k | z_{t}=1, y_{[t-1]}\right) & \propto N_{k, t-1} \\
p\left(y_{t}=\max\left(y_{[t-1]}\right)+1 | z_{t}=1, y_{[t-1]}\right) & \propto \alpha \label{eq:alpha-conditioned}, 
\end{align}
where $N_{k, t-1}$ is the number of blocks of contiguous utterances of speaker $k$. The probability of switching to a new speaker is controlled by the parameter $\alpha$ which is critical for the correct functioning of the whole framework: large values of $\alpha$ force the model to over estimate the number of speakers, instantiating several networks; conversely small values result in limiting the number of speakers by merging clusters.

With respect to the estimation performed in \cite{zhang2019fully}, we propose the following analytical formulation for $\alpha$:
\begin{equation}
\label{eq:crp-alpha}
    \alpha = \frac{\sum_{m = 1}^{|\boldsymbol{\mathcal{D}}|} \left(\max \left( \Y_m\right) - 1\right)}{\sum_{m = 1}^{|\boldsymbol{\mathcal{D}}|}\sum_{t=1}^{|\Y_m|} \mathbbm{1}\left(y_{m,t}\neq y_{m,t+1}\right) }.
\end{equation}
This formulation has the advantage that it can be derived from eq. \ref{eq:alpha-conditioned}, and therefore it is semantically coherent with the role of the parameter. In addition, the value of the parameter is estimated straight from the data, independently of any the error metric or heuristic.

\begin{figure*}[!ht]
\centering
  \includegraphics[width=0.85\textwidth]{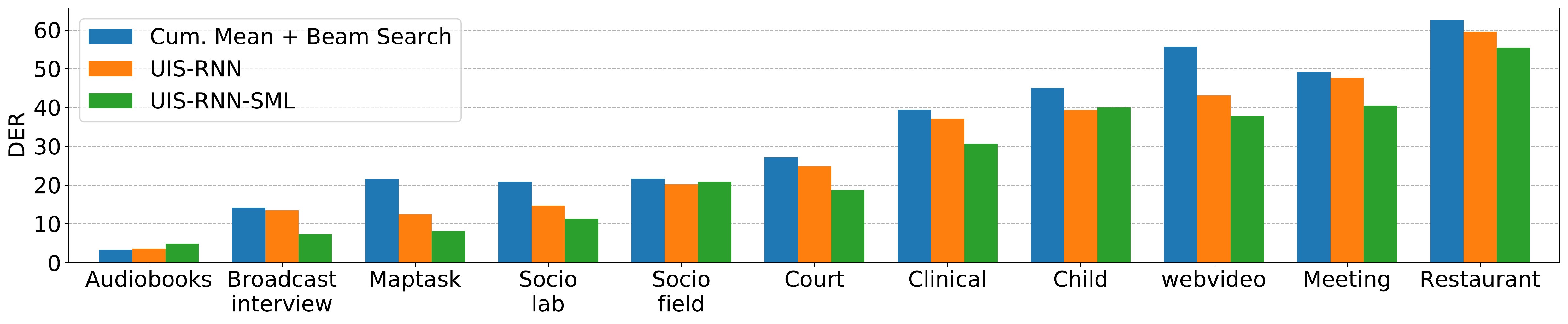}
  \vspace{-0.15cm}
  \caption{DER for each domain in track 1 of the DIHARD-II test set. Domains are displayed in ascending order of difficulty.}
  \vspace{-0.15cm}
\label{fig:disaggDER}
\end{figure*}

\section{Experiments and results}
\label{sec:experiments}

\subsection{Dataset}
\label{sec:dataset}
We train and evaluate our method on the data used in the DIHARD-II challenge~\cite{Ryant2019}. The challenge features two audio input conditions: single channel and multi-channel. We focus on single channel data with reference~\ac{SAD}, as per the track 1 of the competition. The dataset is divided into two subsets, development and evaluation, each consisting of selections of 5-10 minute audio files sampled from 11 different conversational domains for a total of approximately 2 hours of audio.

Using stratified holdout, we further split the development set into training set (80\%) and validation set (20\%). Also, we randomize the holdout procedure, such that for every experiment we get a different data partitioning. Stratification is performed over the set of domains, according to their frequency in the whole development set.

Although the proposed approach does not handle cases where multiple speakers are active simultaneously, we do not exclude overlapping speech segment from the training material. In fact, we observed that considering multi-speaker segments as a separate speaker slightly improves performance. 

\subsection{Experimental setup}
\label{sec:setup}
We use as speaker embeddings of our supervised diarization system x-vectors~\cite{Sell2018} using the pre-trained models available in the Kaldi diarization recipe~\cite{Povey_ASRU2011}. X-vectors with dimension 512 are extracted from non-overlapped 1 second speech segments and are subsequently reduced to dimension 200 with~\ac{PCA} before feeding them to the model.

For what concerns the sequence generation component, our network resembles the architecture presented in~\cite{zhang2019fully}. However, since we are using different features, x-vectors on fixed-length segments instead of d-vectors extracted from ground truth speaker segments, we explored several configurations varying the sizes of the layers. We found that reasonable results are obtained using one recurrent and one fully connected layer with 200 units each.

The other two parameters, $p_0$ and $\alpha$ for the speaker change and the speaker assignment components respectively, are estimated using their analytical formulations. For the transition probability $p_0$ we apply the same formula as in \cite{zhang2019fully}, while for $\alpha$ we use eq.~\ref{eq:crp-alpha}. We also explored some search based techniques for hyperparameter optimization, like grid search and line search, but we found they do not provide noticeable improvements in performance. Furthermore, the value for the variance of the observations $\sigma^2$ is optimized during training using Adam, as in \cite{zhang2019fully}. 

Apart from the SML loss, two more regularization losses help the model to converge~\cite{uisrnn-official-library}. The first one is a simple L2 loss on the parameters of the GRU, the second one uses an inverse gamma distribution to regularize the value of $\sigma^2$ that would otherwise diverge to very large values.

In inference we use beam search with beam size $\beta=15$. Unlike in \cite{zhang2019fully}, in our dataset we can not consider the number of speakers to be bounded. 
This makes inference expensive.

Networks are trained several times using Adam optimizer and the best model is selection by measuring the~\ac{DER} on the validation set, using a smaller beam width ($\beta=2$) to reduce the computational cost. 

~\ac{DER} is measured using \textit{dscore}~\cite{dscore}, the official scoring tool of DIHARD-II competition which does not account for any forgiveness collar, considering also overlapped speech segment. However, since none of the methods under evaluation handle overlapped speech we also report performance without overlap. 


\subsection{Results}
\label{sec:results}
\begin{table}[!ht]
\begin{center}
    \begin{tabular}{ |c|c|c| } 
        \hline
        \textbf{Method} & \textbf{DER} & \textbf{DER - no overlap}\\ 
        \hline
        \hline
        Cum. mean + beam search  & 34.0 & 26.7\\
        UIS-RNN~\cite{zhang2019fully}\cite{uisrnn-official-library} & 30.9 & 23.4\\
        UIS-RNN + eq.\ref{eq:crp-alpha} & 30.3 & 22.8\\ 
        UIS-RNN-SML + eq.\ref{eq:crp-alpha} & \textbf{27.3} & \textbf{19.4}\\
        \hline
        \hline
        PLDA + AHC~\cite{Ryant2019} (offline) & 26.1 & 17.7\\
        \hline
    \end{tabular}
\end{center}
\caption{\ac{DER} on track 1 of DIHARD II test data, with and without overlapping speech. \ac{PLDA}+\ac{AHC} refers to the off-line baseline provided with the challenge. $N=2$ in UIS-RNN-SML.}
\label{tab:DER}
\end{table}

Table~\ref{tab:DER} reports the performance of our proposed UIS-RNN-SML, based on SML and $\alpha$ estimation, in comparison against two online baselines. The first one is a na\"ive implementation in which the GRU is replaced by a simple cumulative mean of the embeddings (Cum. mean + beam search in Table~\ref{tab:DER}). This na\"ive baseline helps highlighting the contribution of the neural network, disentangling it from the other components of the framework. The second is the original UIS-RNN~\cite{zhang2019fully}, using the implementation provided in \cite{uisrnn-official-library}. 
%
%
To give an idea of how difficult the task is, we also report the offline baseline provided in the DIHARD-II challenge \cite{Ryant2019}, which performs diarization by scoring the x-vectors with \ac{PLDA} \cite{sell2014speaker}, and clustering using \ac{AHC} \cite{han2008strategies}.

The na\"ive implemenation based on cumulative mean with beam search is outperformed by the UIS-RNN by a large margin, with and without overlapping speech segments. This confirms that the simple mean of a partial sequence of embeddings does not properly model the speaker and that the neural network makes an active contribution. 
A further small but significative~\ac{DER} reduction, both with and without overlap, with respect to the original implementation is provided by estimating $\alpha$ with eq.~\ref{eq:crp-alpha} (third row in Table~\ref{tab:DER}). 

\begin{figure}[!ht]
\centering
  \includegraphics[width=0.75\columnwidth]{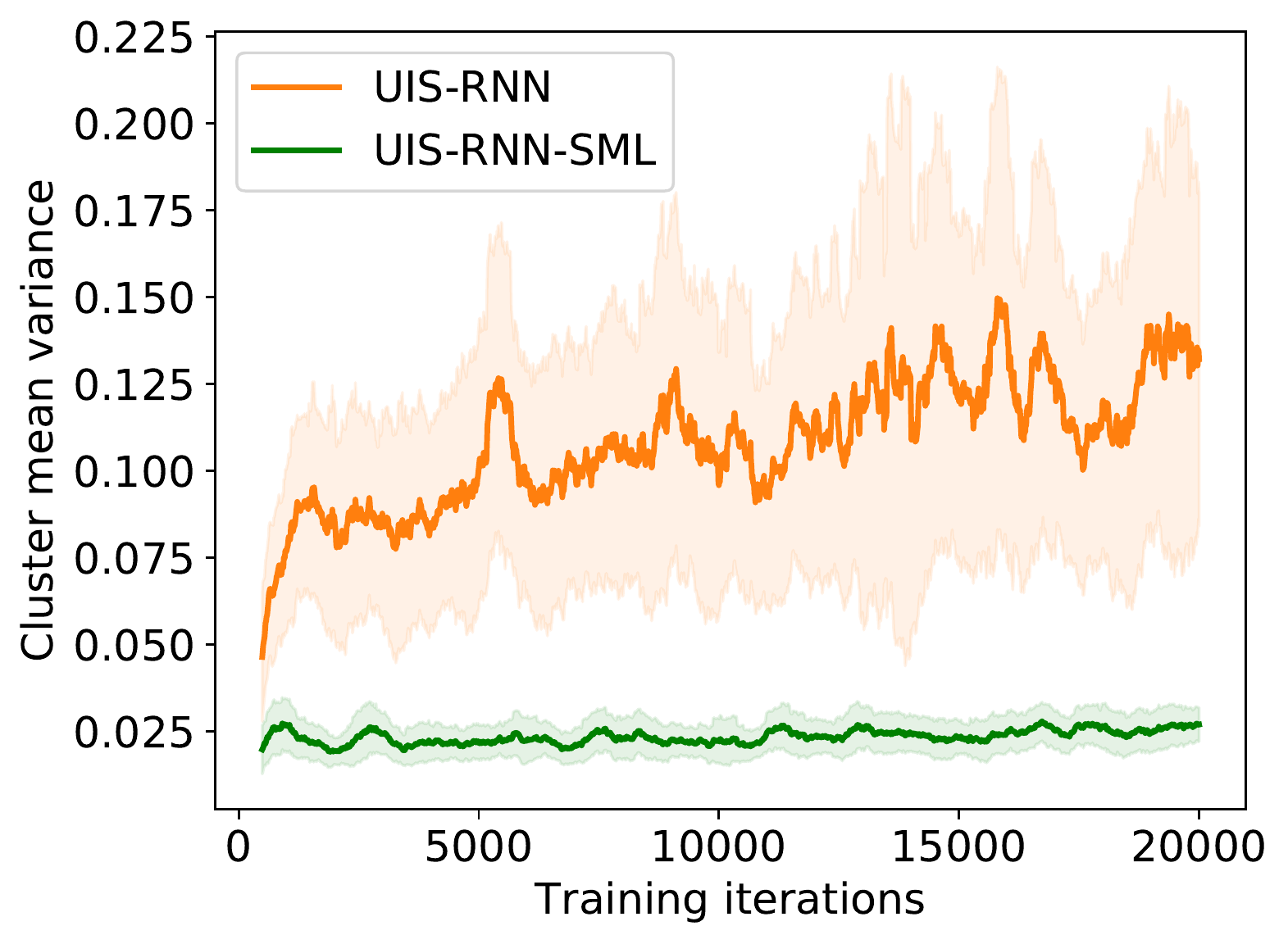}
  \caption{Cluster mean variance over sequences during the first 20K training iterations.}
  \label{fig:tmv}
\end{figure}

Finally, a larger leap in performance is achieved by replacing the original loss function with the \ac{SML} we proposed. Note that the UIS-RNN-SML achieves similar performance to the offline baseline used in DIHARD-II~\cite{Ryant2019}, although online unsupervised clustering algorithms usually perform significantly worse than offline clustering algorithms. The performance improvement is due the regularizing effect introduced by the SML in training. We observed that, keeping \textit{learning rate} and \textit{batch size} fixed, training with SML is much less noisy than the original one: using the more accurate supervision given by the sample mean 
results in better gradients, which in turn helps convergence to deeper minima. The stabilizing effect of the SML is evident in Fig.~\ref{fig:tmv} where we report the variance of the means of the speaker clusters generated by the network during training. Models trained with eq.~\ref{eq:sample-mean} exhibit less output variance compared to those trained with eq.~\ref{eq:networktraining}. This behaviour turns out to be very beneficial in the decoding phase when the means of the clusters should not change dramatically while the sequence unfolds. 

For a better understanding of the behaviour of our proposed method, Fig.~\ref{fig:disaggDER} reports the \ac{DER} for each context in the dataset. Our method is better than the original UIS-RNN in all the most challenging contexts, except for ``socio field'' and ``child'', where our performance is basically aligned to the other methods. We observe a small performance deterioration in ``Audiobooks''. This occurs because the UIS-RNN-SML, predicting the mean more accurately, produces slightly smaller values for the cluster variance $\sigma^2$. Although this is beneficial in most cases, it can marginally reduce performance in contexts with very low number of speakers. This disadvantage can be partially alleviated by defining context dependent $\alpha$ and $p_0$.

\begin{figure}[!ht]
\centering
  \includegraphics[width=0.65\columnwidth]{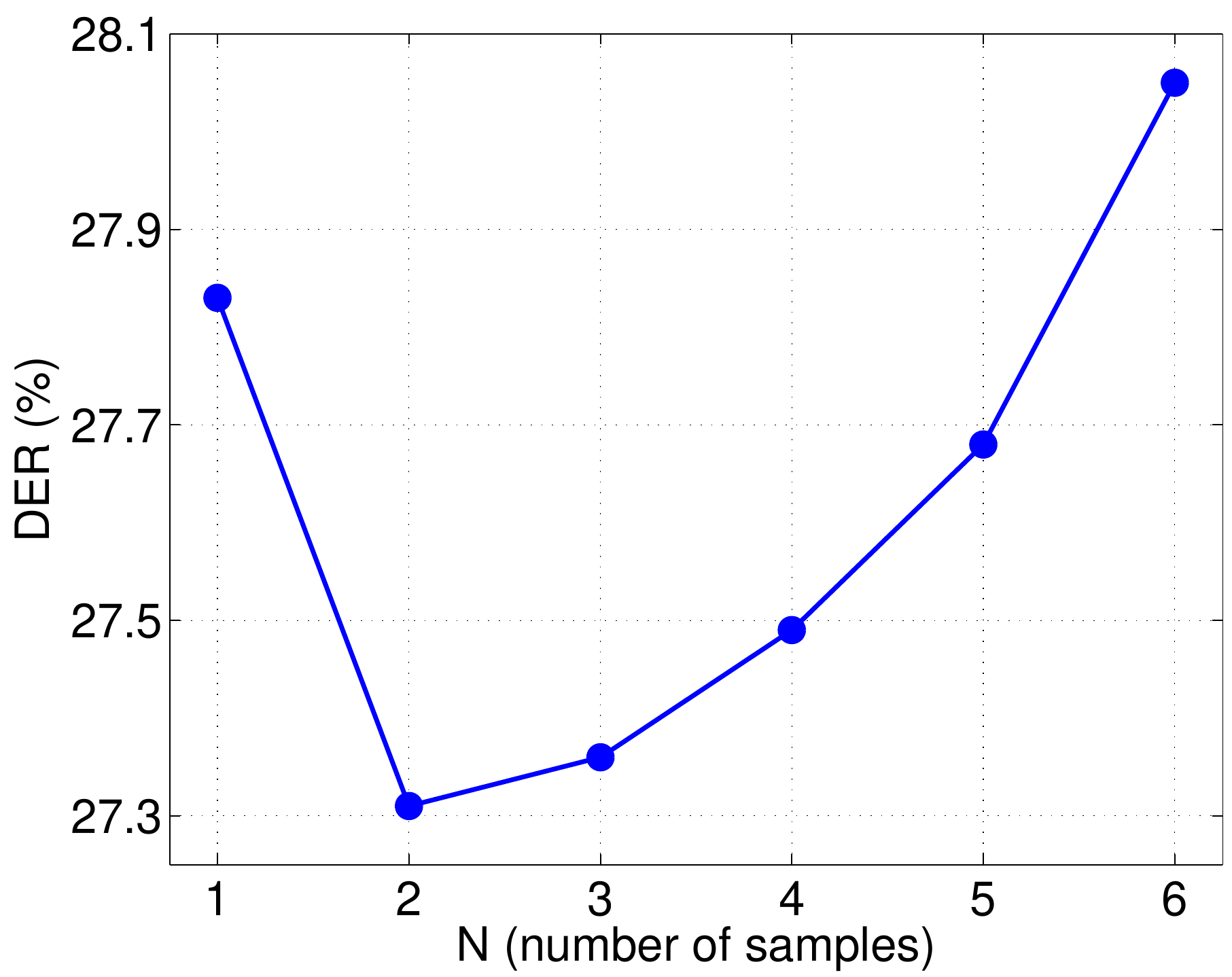}
  \caption{\ac{DER} on track 1 of DIHARD II test data, varying number of samples $N$ in SML.}
  \label{fig:nsamples}
\end{figure}

Finally we evaluate the impact of the number of samples $N$ used to estimate the mean of the distribution. Fig.~\ref{fig:nsamples} shows the~\ac{DER} on the whole evaluation set for different values of $N$. On these data, $N=2$ provides the lower~\ac{DER}, but values from 2 to 4 produce very similar results. Unsurprisingly, performance degrades using larger values for $N$, due to overfitting, because the sample mean approximates the real mean too tightly. Note that the case $N=1$ would be equivalent to the UIS-RNN except for the fact that observations are sampled with replacement. This gives a considerable improvement (27.83\% against 30.3\%) because outliers of the speaker clusters are less likely to be observed by the network as targets during training, reducing the overall variance of the output. 
\section{Conclusions}
\label{sec:conclusions}
In this paper we presented an evolution of a supervised speaker diarization system where the clustering module is replaced by a trainable model called unbounded interleaved-state RNN. Specifically, we proposed a modified loss function that stimulates the neural network to model speakers more accurately. In addition, we introduced a semantically grounded formulation for the estimation of the parameter that controls the speaker assignment probability. We evaluated the proposed online diariaztion approach on the DIHARD-II multi-domain data, showing, through extensive experiments, that it outperforms the original UIS-RNN formulation. Finally, we fully disclose our code and trained models to make our results reproducible.

\newpage
\bibliographystyle{IEEEbib}
\bibliography{refs}

\end{document}